\documentclass[letterpaper, 10 pt, conference]{ieeeconf}  

\IEEEoverridecommandlockouts                              

\usepackage{graphicx} 
\usepackage{amsmath} 
\usepackage{todonotes}
\usepackage{cite}
\usepackage{tikz, pgfplots}
\usepackage{float}

\makeatletter
\let\NAT@parse\undefined
\makeatother
\usepackage{hyperref}

\usetikzlibrary{
	pgfplots.colorbrewer,
}
\pgfplotsset{
	cycle list/.define={my marks}{
		every mark/.append style={solid,fill=\pgfkeysvalueof{/pgfplots/mark list fill}},mark=*\\
		every mark/.append style={solid,fill=\pgfkeysvalueof{/pgfplots/mark list fill}},mark=square*\\
		every mark/.append style={solid,fill=\pgfkeysvalueof{/pgfplots/mark list fill}},mark=triangle*\\
		every mark/.append style={solid,fill=\pgfkeysvalueof{/pgfplots/mark list fill}},mark=diamond*\\
	},
}

\title{\LARGE \bf
Reducing Uncertainty by Fusing Dynamic Occupancy Grid Maps in a Cloud-based Collective Environment Model*
}

\author{Bastian Lampe$^{1}$ and Raphael van Kempen$^{1}$, Timo Woopen$^{1}$,\\ Alexandru Kampmann$^{2}$, Bassam Alrifaee$^{2}$,  Lutz Eckstein$^{1}$
\thanks{*This research is accomplished within the project "UNICAR\textit{agil}" (FKZ 16EMO0289). We acknowledge the financial support for the project by the Federal Ministry of	Education and Research of Germany (BMBF).}
\thanks{$^{1}$The authors are with the \mbox{Institute for Automotive Engineering (ika)}, 
	RWTH Aachen	University, 52074 Aachen, Germany
	{\tt\small \{firstname.lastname\}@ika.rwth-aachen.de}}%
\thanks{$^{2}$The authors are with \mbox{Informatik 11 - Embedded Software (i11)},
	RWTH Aachen University,	52074 Aachen, Germany
	{\tt\small \{lastname\}@embedded.rwth-aachen.de}}%
}

\newcommand\copyrighttext{%
	\footnotesize \copyright{ }2020 IEEE. Personal use of this material is permitted. Permission from IEEE must be obtained for all other uses, in any current or future media, including reprinting/republishing this material for advertising or promotional purposes, creating new collective works, for resale or redistribution to servers or lists, or reuse of any copyrighted component of this work in other works.}
\newcommand\copyrightnotice{%
	\begin{tikzpicture}[remember picture,overlay]
	\node[anchor=south,yshift=10pt] at (current page.south) {\parbox{\dimexpr\textwidth-\fboxsep-\fboxrule\relax}{\copyrighttext}};
	\end{tikzpicture}%
}

\begin{document}

\maketitle
\thispagestyle{empty}
\pagestyle{empty}

\copyrightnotice

\begin{abstract}

Accurate environment perception is essential for automated vehicles. Since occlusions and inaccuracies regularly occur, the exchange and combination of perception data of multiple vehicles seems promising. This paper describes a method to combine perception data of automated and connected vehicles in the form of evidential Dynamic Occupany Grid Maps (DOGMas) in a cloud-based system. This system is called the Collective Environment Model and is part of the cloud system developed in the project UNICARagil. The presented concept extends existing approaches that fuse evidential grid maps representing static environments of a single vehicle to evidential grid maps computed by multiple vehicles in dynamic environments. The developed fusion process additionally incorporates self-reported data provided by connected vehicles instead of only relying on perception data. We show that the uncertainty in a DOGMa described by Shannon entropy as well as the uncertainty described by a non-specificity measure can be reduced. This enables automated and connected vehicles to behave in ways not before possible due to unknown but relevant information about the environment. 

\end{abstract}

\section{INTRODUCTION}

Three major goals are often stated when developing automated vehicles~(AVs). Transportation shall become more safe, more efficient and more convenient. In order to reach all three goals, it is essential for an AV's behavior and trajectory planning to have access to an accurate representation of the vehicle's environment. In the project UNICARagil~\cite{Woopen2018}, a modular service-oriented software architecture called "Automotive Service-Oriented Architecture" (ASOA)~\cite{Kampmann2019} is developed. It supports these goals by orchestrating all services available to a vehicle into one driving function based on the quality of the data, which the different services may provide. Importantly, not all of the services need to be located in the vehicle. The Collective Environment Model~(CEM), part of which is described in this paper, runs on a cloud server to which vehicles are connected. A more detailed overview of the cloud services developed in UNICARagil can be found in \cite{Lampe2019}. The CEM provides additional data to vehicles that aims at improving their individual perception. There are three main services provided by the CEM.

\begin{enumerate}
	\item Provision of a fused environment representation computed from those of multiple individual vehicles,
	\item Assessment of the accuracy of the environment representations of individual vehicles by evaluating their consistency,
	\item Provision of estimates of the future state of a vehicle's environment.
\end{enumerate}

The presented research focuses on the description and evaluation of the first aspect. 

\section{Contribution}

We describe and evaluate a mechanism that combines the evidential Dynamic Occupancy Grid Maps~(DOGMas) of multiple vehicles connected to the CEM. The architecture of the CEM is similar to a conventional environment model architecture for automated vehicles, as described in e.g.~\cite{Winner2015}. New aspects most importantly include its execution on a cloud server and the extended functionalities that are only possible through the fusion of data from multiple vehicles instead of only from multiple sensors. Since data is transmitted to the CEM wirelessly, challenges concerning the latency between sensor measurements and the availability of a computed environment representation in a vehicle arise. In contrast, opportunities originate from the possibility to incorporate the self-reported data that connected vehicles may share. The perception and prediction algorithms can benefit from additional data which may contain a vehicle's pose, dimensions, motion state and planned trajectories. The CEM can fuse different types of environment representations such as sensor data, object lists, DOGMas and free space representations. In this paper, we focus on the fusion of evidential DOGMas. We extend the fusion algorithms described in \cite{Thrun2006, Camarda2018, Dezert2015, Grimmer2017} such that they are capable of fusing non-synchronized evidential occupancy grids of multiple connected vehicles that describe dynamic environments. The self-reported motion states, poses and dimensions of vehicles connected to the CEM are incorporated in the developed process to further reduce uncertainty contained in the computed DOGMas. The latency as part of the Quality of Service~(QoS) is constantly assessed and considered in the developed mechanism.

\section{Background}

Before describing the mechanism and applying it to an example scenario, let us explore existing research that provides the foundation for our work. Occupancy Grid Maps are first outlined and then extended by principles of evidence theory. This allows us to specifically quantify uncertainty resulting from a lack of evidence in an Occupancy Grid Map. Then, the motivation and mechanisms regarding DOGMas are briefly elaborated. They provide the necessary input to the cloud-based fusion mechanism for evidential DOGMas of multiple vehicles developed in our research.

\subsection{Occupancy Grid Maps}

Occupancy Grid Maps as described in \cite{Thrun2006} and \cite{Elfes1989} represent the environment of a vehicle as an evenly spaced grid where each grid cell contains a binary random variable corresponding to the occupancy state at the location of that grid cell. For a vehicle that is not connected to other vehicles, the posterior estimate of a cell state uses the vehicle's sensor measurements. To reduce uncertainty, an occupancy grid mapping algorithm may combine the measurements of multiple sensors. Often, a binary Bayes filter is used in combination with an inverse sensor model to combine the information contained in a measurement of a lidar sensor as described in  \cite{Winner2015},\cite{Thrun2006},\cite{Nuss2016}. The binary occupancy state $ o_k \in \{O, F\} $ ($O$: ~Occupied; $F$: ~Free) at time $k$ is represented as an occupancy probability $ p_{k}(o_k) $ in the cells. An inverse sensor model computes an occupancy probability $ p_{\boldsymbol{z}_{k+1}}(o_{k+1}|\boldsymbol{z}_{k+1}) $ from a measurement $ \boldsymbol{z}_{k+1} $ at time $k+1$. The last estimated state of the grid cells and the occupancy probability computed from the current measurement can be combined by the binary Bayes filter \cite{Thrun2006}, \cite{Nuss2016}.

The binary Bayes filter relies on the Markov assumption which states that the occupancy state of a cell is independent of past measurements \cite{Thrun2006}. Additionally, the true state of a cell must not change over time \cite{Nuss2016}. All of these restrictions limit the usefulness of the binary Bayes filter in a dynamically changing environment as experienced in automated driving.  Another weakness of the binary Bayes filter constitutes in the fact that for an uncertain cell state ($p_{k}(o_k) \approx 0.5$), we cannot differentiate between the underlying reasons for the uncertainty. It can either result from an absence of evidence for the occupancy state of a cell or from uncertain, potentially conflicting evidence.

\subsection{Evidence Theory}

Evidence Theory as described by Dempster and Shafer (DST) \cite{Shafer1976} can be used to establish an extended representation of the state of a cell and provide a mechanism to combine the current estimate of a cell state with new evidence, based on a more recent measurement, to form a new estimate. It also allows the differentiation between uncertain and unknown states.

In the case of occupancy grids, the so called frame of discernment (FOD) $ \Theta = \{ F, O \} $ represents a set of mutually exclusive possible cell states Occupied ($O$) and Free ($F$). The power set
\begin{equation}
\begin{aligned}
\label{eq:ds_powerset}
2^\Theta = &\{ \emptyset, \{F\}, \{O\}, \Theta\ \}.
\end{aligned}
\end{equation}
contains all possible subsets of $\Theta$. Each element of $2^\Theta$ is associated with a belief mass. In our case, the belief mass of the element $ \Theta $ is a measure of the lack of evidence for the singleton hypotheses of $2^\Theta$. A basic belief function $m$ assigns a belief mass to each element of $ \Theta $ where the mass of the empty set is zero and the sum of all other masses equals one.
\begin{align}
	m: 2^\Theta &\rightarrow [0,1] \label{eq:dempster_mass_zero_one}\\
	m(\emptyset) &= 0  \\
	\sum_{A \in 2^\Theta} m(A) &= 1 \label{eq:dempster_sum_masses_one}
\end{align}

By using DST, the classic probability for the state of a cell $p(X)$ is bounded by the belief $ bel(X) $ and the plausibility $ pl(X) $ for a state of a cell. Both belief and plausibility for a specific state of a cell can be computed from the masses of the elements of $ 2^\Theta $ for that state. The belief for a hypothesis $X$ results from the sum of all masses associated with elements $ A $ of $ 2^\Theta $ that are a subset of $X$. The plausibility is the sum of those elements $ A $ of $ 2^\Theta $ that contain $X$ as a subset.

\begin{align}
	bel(X) &= &\sum_{A \in 2^\Theta | A \subseteq X} m(A)& \\
	pl(X)  &= &\sum_{A \in 2^\Theta | X \subseteq A} m(A)& = 1-bel(\neg X)
\end{align} 
We can use the pignistic transform to get the classic, one-dimensional probability from the masses of $ 2^\Theta $ \cite{Nuss2016}. For $ 2^\Theta = \{ \emptyset, \{F\}, \{O\}, \Theta\} $, we get 

\begin{eqnarray}
\label{eq:pignistic_tf}
p_O = p(o_k=O) = m_k(O) + 0.5 \cdot (1 - m_k(O) - m_k(F)).
\end{eqnarray}
When combining the evidence from multiple sources such as multiple sensors or, as in our case, multiple vehicles, we can apply Dempster's Rule of Combination. It gives us the joint mass $m_{12}(X)$ for a hypothesis $ X $ which is an element in the power sets of the combinable FODs $ 2^{{\Theta}_1} $ and $ 2^{{\Theta}_2} $ of the different sources. 

\begin{equation}
\label{eq:dempsters_rule_of_combination}
m_{12}(X) = (m_1 \oplus m_2)(X) = 
\frac{(m_1 \cap m_2)(X)}
{1 - K}
\end{equation}
The rule normalizes the agreeing evidence 

\begin{equation}
	(m_1 \cap m_2)(X) = \sum\limits_{A,B \in 2^\Theta | A \cap B = X} m_1(A) \cdot m_2(B)
\end{equation}
by the factor $1 - K$, where 

\begin{equation}
	 K=\sum\limits_{A,B \in 2^\Theta | A \cap B = \emptyset} m_1(A) \cdot m_2(B)
\end{equation}
is a measure for the conflict between the beliefs based on the two sources.

\subsection{Measures of Uncertainty}\label{sec:measures_of_uncertainty}

We base our uncertainty quantification on two of the dimensions of uncertainty described in \cite{Reineking2014}. By using the pignistic transform, we are able to compute the uncertainty of a cell in a DOGMa described by Shannon entropy 

\begin{equation}
H(p)=-\sum_{x\in \Theta}p_O(x)\log_2 p_O(x).
\end{equation}
A possible non-specificity measure which quantifies uncertainty resulting from a lack of evidence is defined by \cite{Dubois1985} as

\begin{equation}
NS(m) = \sum_{X\in \Theta , X\neq \emptyset}m(X)\log_2 |X|.
\end{equation}
For an Occupancy Grid Map with $\Theta=\{O,F\}$, this simplifies to

\begin{equation}
NS(m)=m(\Theta),
\end{equation}
which will serve as our second evaluation metric. For the arithmetic mean of $H(p)$ and $NS(m)$ over all cells of a DOGMa $D$, we write $\overline{H}\big(D\big)$ and $\overline{NS}\big(D\big)$.

\subsection{Dynamic Occupancy Grid Maps}

A Dynamic Occupancy Grid Map is an extension of a regular Occupancy Grid Map that also captures the dynamics of the environment. An evidential DOGMa that is based on DST may additionally allow to differentiate between uncertain and unknown states. As described in \cite{Nuss2016}, the state of a cell can be represented as a vector $ \Omega $ where

\begin{equation}
	\Omega = [m_O, m_F, v_N, v_E, \sigma_{v_N}^2,	\sigma_{v_E}^2,	\sigma_{v_N v_E}]^T
\end{equation}
The variables $ v_N $ and $ v_E $ describe the velocity of an object at the location of the cell in north and east direction. The variances $ \sigma_{v_N}^2 $, $ \sigma_{v_E}^2 $ and the covariance $ \sigma_{v_N v_E} $ may also be provided by the algorithm computing the DOGMa. The mass function associated with the FOD is not explicitly contained in the DOGMa because it can be obtained using equation \ref{eq:dempster_sum_masses_one}. In this paper, we do not develop the DOGMa algorithm itself. The input format of the developed fusion process in our work is based on the possible results provided by the algorithm described in \cite{Nuss2016}, where the DOGMa is interpreted as a Random Finite Set. They use a Probability-Hypothesis-Density filter that is approximated by multiple instances of a Bernoulli filter where the state estimation is discretized by a particle approach. \\

The use of DOGMas enables us to incorporate the representation of the dynamics of the environment in the fusion algorithm. This can be helpful when dealing with Occupancy Grid Maps that contain moving objects and that are not synchronized, as expected in the cloud use case described in this work.

\section{Cloud-Based DOGMa Fusion}

The proposed mechanism involves several steps that  differentiate between several different types of DOGMas.

\begin{enumerate}
	\item Individual vehicles $i$ compute local (superscript ${(l)}$) DOGMas $D^{(l_i)}$ of size $d^{(l)} \times d^{(l)} \times f$ where $f$ denotes the number of features in a cell and $d^{(l)}$ the number of cells in a row and a column of the $D^{(l)}$ respectively.
	\item The CEM contains a collective (superscript ${(c)}$) DOGMa $D^{(c)}$ of size $d^{(c)} \times d^{(c)} \times f$ where $d^{(c)}$ denotes the number of cells in a row and a column of the $D^{(c)}$ respectively. 
	\item $D^{(c)}$ is estimated in a process involving a prediction of $\hat{D}^{(c)}_{k|k-1}$ and an updated version $\hat{D}^{(c)}_{k|k}$
	\item The CEM extracts sub maps $\hat{D}^{(l_i)}_{k}$ from $\hat{D}^{(c)}_{k}$ and performs a prediction which yields $\hat{D}^{(l_i)}_{k+1|k}$ that can be used by individual vehicles.
\end{enumerate}
In this work, we choose the cell side length $c$ to be $0.15$~$cm$, $d^{(l)}=667$ and $d^{(c)} = 1000$. \\

\noindent The process consists of the following steps: \\

\begin{enumerate}
	\item Define a traffic area for which $D^{(c)}$ shall be computed.\label{item:traffic_area}
	\item Periodically receive poses of vehicles.
	\item Periodically assess the transmission latency.
	\item Detect whether a $D^{(l_i)}$ and a $D^{(c)}$ overlap.
	\item Initialize $D^{(c)}$ if there is an overlap.
	\item Remove $D^{(c)}$ if there is no overlap.
	\item Receive a new $D^{(l_i)}_{k}$.
	\item Predict $\hat{D}^{(c)}_{k|k-1}$ from $\hat{D}^{(c)}_{k-1|k-1}$
	\item Fuse $\hat{D}^{(c)}_{k|k-1}$ and $D^{(l_i)}_{k}$ to yield $\hat{D}^{(c)}_{k|k}$.
	\item Extract $\hat{D}^{(l_i)}_{k|k}$ from $\hat{D}^{(c)}_{k|k}$ for each vehicle such that poses and sizes of $\hat{D}^{(l_i)}_{k|k}$ conform with those of $D^{(l_i)}_{k+1}$ of receiving vehicles.
	\item Compute all $\hat{D}^{(l_i)}_{k+1|k}$ from corresponding $\hat{D}^{(l_i)}_{k|k}$.
	\item Transmit each $\hat{D}^{(l_i)}_{k+1|k}$ to vehicle $i$.
\end{enumerate}

\begin{figure}[t]
	\centering
	\includegraphics[trim= 40 60 60 60, clip, width=3in]{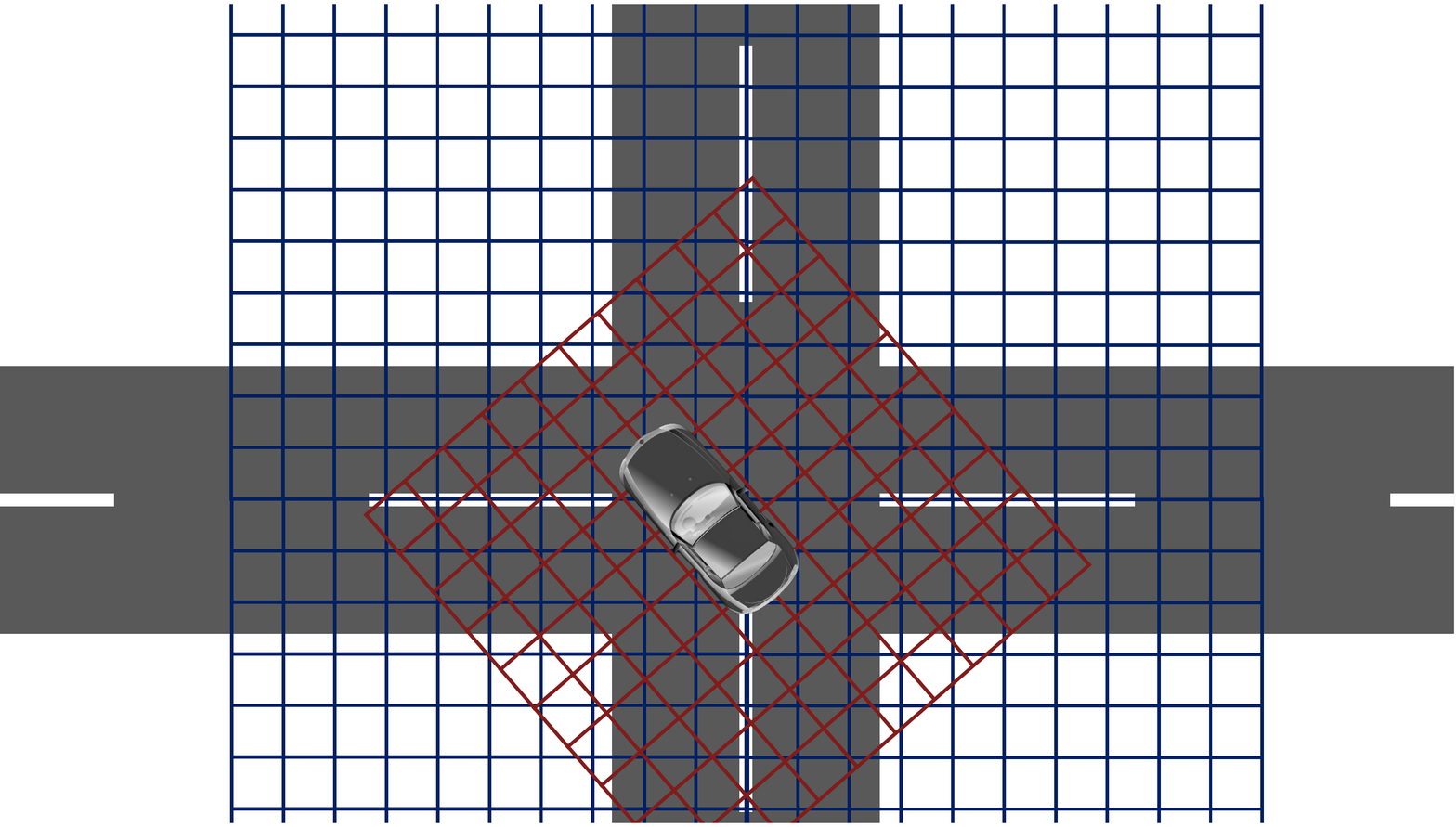}
	\caption{A $D^{(c)}$ (blue) is located at a fixed position and fixed orientation in the world. A $D^{(l_i)}$ (red) is always at a fixed position and orientation relative to the vehicle $i$ that provides the $D^{(l_i)}$ to the CEM}
	\label{fig:traffic area}
\end{figure}

\noindent A traffic area is a defined rectangular region for which a $D^{(c)}$ may be initialized. The latter does not necessarily align with the $D^{(l_i)}$ from which it is computed, which can be seen in Fig. \ref{fig:traffic area}.
Each vehicle connected to the cloud periodically transmits its pose in the ETRS89 coordinate frame. This later allows the allocation of the cells of a $D^{(l)}$ to the cells of the $D^{(c)}$ and vice versa. There needs to be an overlap between two $D^{(l_i)}$ and the traffic area but not necessarily between two $D^{(l_i)}$ in order for the initialization of the $D^{(c)}$ to take place. Since information of non-overlapping parts of the $D^{(l_i)}$ persists in $D^{(c)}$ for a certain time, a vehicle may benefit from parts of the $D^{(c)}$ that currently lie beyond the boundaries of its $D^{(l_i)}$. When overlap is detected for the first time, all cells of a $D^{(c)}$ of the respective traffic area are initialized with the belief masses $m_O=m_F=0$ corresponding to an unknown state. Every time a $D^{(l_i)}$ is received, its creation timestamp and the timestamp of its arrival are stored in order to later assess the current latency between the vehicles and the cloud.

\begin{figure}[t]
	\centering
	\includegraphics[trim=244 0 275 170,clip, width=3in]{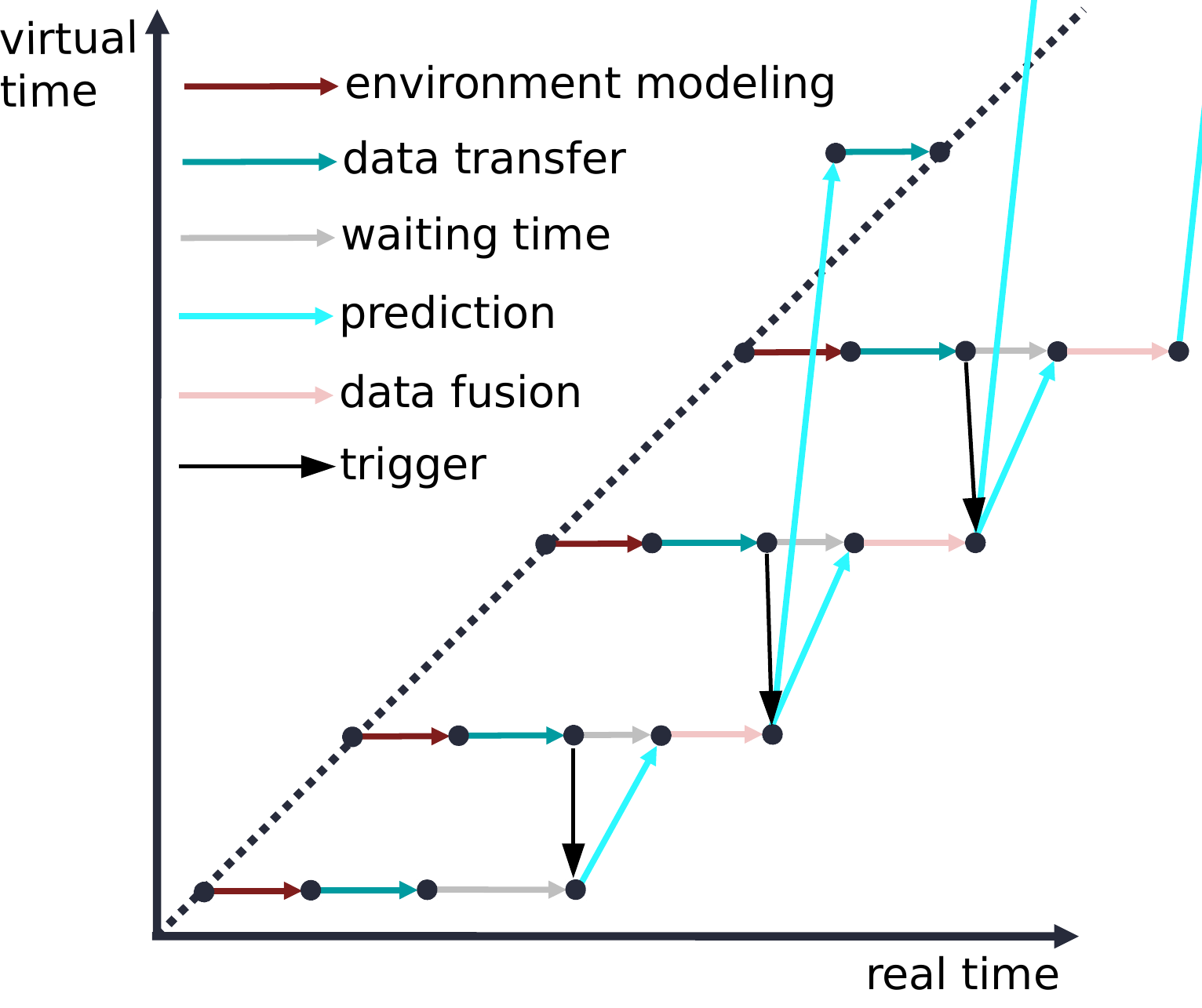}
	\caption{The output of the vehicles' environment modeling algorithm is transmitted to the CEM. As soon as a second environment representation is received, a trigger containing the (real) time stamp of the environment representation initiates a prediction of the current environment representation in the CEM to that time stamp. Now, the synchronized data may be fused. The updated environment representation may then be predicted to a virtual time that equals real time when the data is received in a vehicle. The prediction horizon is based on the currently measured transmission latency between cloud and vehicle.}
	\label{fig:sync}
\end{figure}

We want the $\hat{D}^{(l_i)}_{k+1|k}$ that are provided by the CEM to represent a vehicle's environment at the same time as it arrives in the vehicle. For that purpose, we continuously assess the latency between vehicles and the cloud. We introduce the concept of ~virtual ~time in order to better visualize the fusion and prediction process. Virtual time is the time that is associated with e.g. an environment representation. Virtual time can be ahead of, equal to and behind real time, i.e. an environment representation can describe the past, current or future environment of a vehicle. As long as no prediction of the future is involved, the virtual time of an environment representation is always behind real time. Fig. \ref{fig:sync} shows one possible fusion scheme that can be used to achieve the aforementioned goal. The necessary fusion scheme is dependent on the processing latency, the transmission latency and the number of connected vehicles. 

To fuse the $\hat{D}^{(c)}_{k-1|k-1}$ with the most recently received $D^{(l_i)}_{k}$, the former is predicted to the timestamp $k$ of $D^{(l_i)}_{k}$. The prediction is based on a model of the processes describing the dynamics of the states of the cells. Multiple different process models are possible:

\begin{itemize}
	\item White-box model: Mathematical model that explicitly uses a priori information about the objects located in the cells and is derived from first principles. 
	\item Black-box model: Mathematical model that makes no a~priori assumptions about the objects located in the cells. The model may be derived from data through machine learning.
	\item Gray-box model: Combination of a white-box and a black-box model.
	\item Cooperation-based: The cell states are predicted based on the planned behavior of traffic participants occupying these cells. This becomes possible when the planned behavior of a traffic participant is communicated to the CEM.
\end{itemize}

In this paper, we use a simple white-box model that assumes linear cell dynamics. It shall act as a base line for future works. The cells of a $D^{(l_i)}_{k}$ received by the CEM are treated as a measurement $\boldsymbol{z}_k$. The estimated state $\boldsymbol{\hat{x}}_{k-1|k-1}$ of each cell of $\hat{D}^{(c)}_{k-1|k-1}$ is propagated through time by predicting the state vector $\boldsymbol{\hat{x}}_{k|k-1} = \left[ x\; y\; v_{x}\; v_{y} \right]^T$. It can be viewed as an estimate of the location of the center of a cell and the velocities assigned to that cell at time $k$ based on information available at time $k-1$. We distribute the belief masses according to the a priori covariance matrix to compute a predicted DOGMa such as $\hat{D}^{(c)}_{k|k-1}$ or $\hat{D}^{(l_i)}_{k+1|k}$.

The estimation and prediction of $\boldsymbol{\hat{x}}$ at time $k$ may be done recursively in a linear quadratic estimation process involving the state covariance matrix $\boldsymbol{P}$, a state transition matrix $\boldsymbol{F}$, the covariance of the process noise $\boldsymbol{Q}$ and the covariance of the observation noise $\boldsymbol{R}$. We can describe the process by a prediction step 

\begin{eqnarray}
\boldsymbol{\hat{x}}_{k|k-1} &=& \boldsymbol{F}_{k} \boldsymbol{\hat{x}}_{k-1|k-1} \\
\boldsymbol{P}_{k|k-1} &=& \boldsymbol{F}_{k} \boldsymbol{P}_{k-1|k-1} \boldsymbol{F}_{k}^T + \boldsymbol{Q}_{k}.
\end{eqnarray} 

that is followed by a correction step 

\begin{eqnarray}
\boldsymbol{\hat{x}}_{k|k} &=& \boldsymbol{\hat{x}}_{k|k-1} + \boldsymbol{K}_k (\boldsymbol{z}_k - \boldsymbol{\hat{x}}_{k|k-1}) \\ 
\boldsymbol{K}_k &=& \boldsymbol{P}_{k|k-1} (\boldsymbol{P}_{k|k-1} + \boldsymbol{R}_k)^{-1} \\
\boldsymbol{P}_k &=& (\boldsymbol{I} - \boldsymbol{K}_k) \boldsymbol{P}_{k|k-1}.
\end{eqnarray}

By applying the transition matrix $\boldsymbol{F}_{k}$ to the state vector of a cell $\boldsymbol{\hat{x}}_{k-1|k-1}$, we get the predicted a priori estimate of the state vector at time $k$. The process noise is largely associated with our assumption of constant velocities. The introduced errors are assumed to be normally distributed here. The use of different process noise models may be examined in later works. For the observation noise, we can use the variances contained in a $\hat{D}^{(l_i)}$ and we can make an assumption based on a model-based maximum acceleration of a vehicle. The latter is reasonable if we do not want to reduce the mean error of our estimation over all scenarios but to reduce the error for scenarios in which vehicles abruptly accelerate or decelerate. If we assume constant linear acceleration, we get the probability $p\in(0,1)$ for the actual state to be inside an interval $I=[\mu-z\sigma;\mu+z\sigma]$ for 

\begin{align}
	&z\sigma_x = z\sigma_y =0.5 \cdot a_{max} \Delta T^2 \\
	&z\sigma_{v_{x}}= z\sigma_{v_{y}} = a_{max} \Delta T
\end{align}
by computing the z-score

\begin{equation}
	z=\Phi^{-1}\Big(\frac{p+1}{2}\Big)=\sqrt{2} \mathrm{erf}^{-1}(p) \label{eq:inv_normal_distribution}
\end{equation}
and choosing the respective $\sigma$ accordingly. 
For a worst case assumption, the parameter $a_{max}$ may for example be chosen such that under Coulomb friction and a coefficient of friction $\mu_f\leq1$, we get a maximum acceleration of $a_{max}=g$. We can then use $p$ as a tuning parameter for how accelerations of vehicles are distributed. The larger we choose $p$ to be, the bigger the assumed noise.\\

Under the previously mentioned assumptions, the process noise matrix can be written as

\begin{align}
	\boldsymbol{Q}_k = \left[\begin{array}{cccc}
	\frac{\Delta T^2}{4} & 0 & 0 & 0 \\
	0 & \frac{\Delta T^2}{4} & 0 & 0 \\
	0 & 0 & 1 & 0 \\
	0 & 0 & 0 & 1		
	\end{array}\right] \frac{\Delta T^2 {}a_{max}^2}{z^2} 
\end{align}

For the observation noise matrix, we apply the same idea. Here, we assume that the major source of uncertainty comes from the discretization in the occupancy grid. In this work, we assume a correct pose estimate. We approximate the probability density function with that of a normal distribution. We assume a location measurement to be located within a symmetric interval around the center of a cell with side length $\Delta c$. The location is assumed to be inside the estimated cell with probability $p$. Again, we apply equation~\ref{eq:inv_normal_distribution} to compute the z-score $z$ and choose $\sigma_x$ and $\sigma_y$ such that the condition $z\sigma_x=z\sigma_y=\frac{\Delta c}{2}$ is satisfied. This gives us the covariance matrix of the observation noise

\begin{align}
\boldsymbol{R}_k &= 
\left[\begin{array}{cccc}
\frac{\Delta{}c^2}{4z^2} & 0 & 0 & 0 \\
0 & \frac{\Delta{}c^2}{4z^2} & 0 & 0 \\
0 & 0 & \sigma^2_{v_x} & \sigma_{v_{xy}} \\
0 & 0 & \sigma_{v_{xy}} & \sigma^2_{v_y}
\end{array}\right].
\end{align}
For the observation noise associated with the cell velocities, we take the variances and covariances contained in $D^{(l_i)}_{k}$.

We can now take the estimate of the location of a cell $\boldsymbol{\mu_x}= \left[x_k, y_k\right]^T$ and distribute the belief masses based on the state covariance matrix. 

The estimated cell location is described by a two-dimensional normal distribution $X \sim \mathcal{N}_2(\boldsymbol{\mu_k}, \Sigma_k)$. For each cell, we can get a probability density function over the coordinates $(x, y)$. We integrate over a cell with center coordinates $(x_c, y_c)$ and side length $\Delta c$ to get the probability of an object occupying a cell that is associated with a belief mass having moved there.

The elements of the covariance matrix $\Sigma_k$ can be extracted from $\boldsymbol{P}_{k|k-1}$ to compute 
\begin{align}
p_c &= \int\limits_{x_{min}}^{x_{max}} ~\int\limits_{y_{min}}^{y_{max}} \frac{1}{2\pi{}\sqrt{|{\Sigma_k}|}} e^{-\frac{1}{2} \left(\left[\begin{smallmatrix}x\\y\end{smallmatrix}\right] - \boldsymbol{\mu_k}\right)^T {\Sigma_k}^{-1} \left(\left[\begin{smallmatrix}x\\y\end{smallmatrix}\right] - \boldsymbol{\mu_k}\right)} dy \, dx
\end{align}
where the integration intervals are defined by the cell boundaries. We may approximate the probability by only evaluating the distribution function at the center of a cell. 

\begin{align}
p_c &\approx
\frac{1}{2\pi{}\sqrt{|\Sigma_k|}} e^{-\frac{1}{2} \left(\left[\begin{smallmatrix}x_c\\y_c\end{smallmatrix}\right] - \boldsymbol{\mu_k}\right)^T {\Sigma_k}^{-1} \left(\left[\begin{smallmatrix}x_c\\y_c\end{smallmatrix}\right] - \boldsymbol{\mu_k}\right)} \cdot \Delta{}c^2 \label{eq:konzept_cem_verteilung_diskret}
\end{align}

\begin{figure}[t]
	\centering
	\includegraphics[trim=220 280 220 70,clip, width=3in]{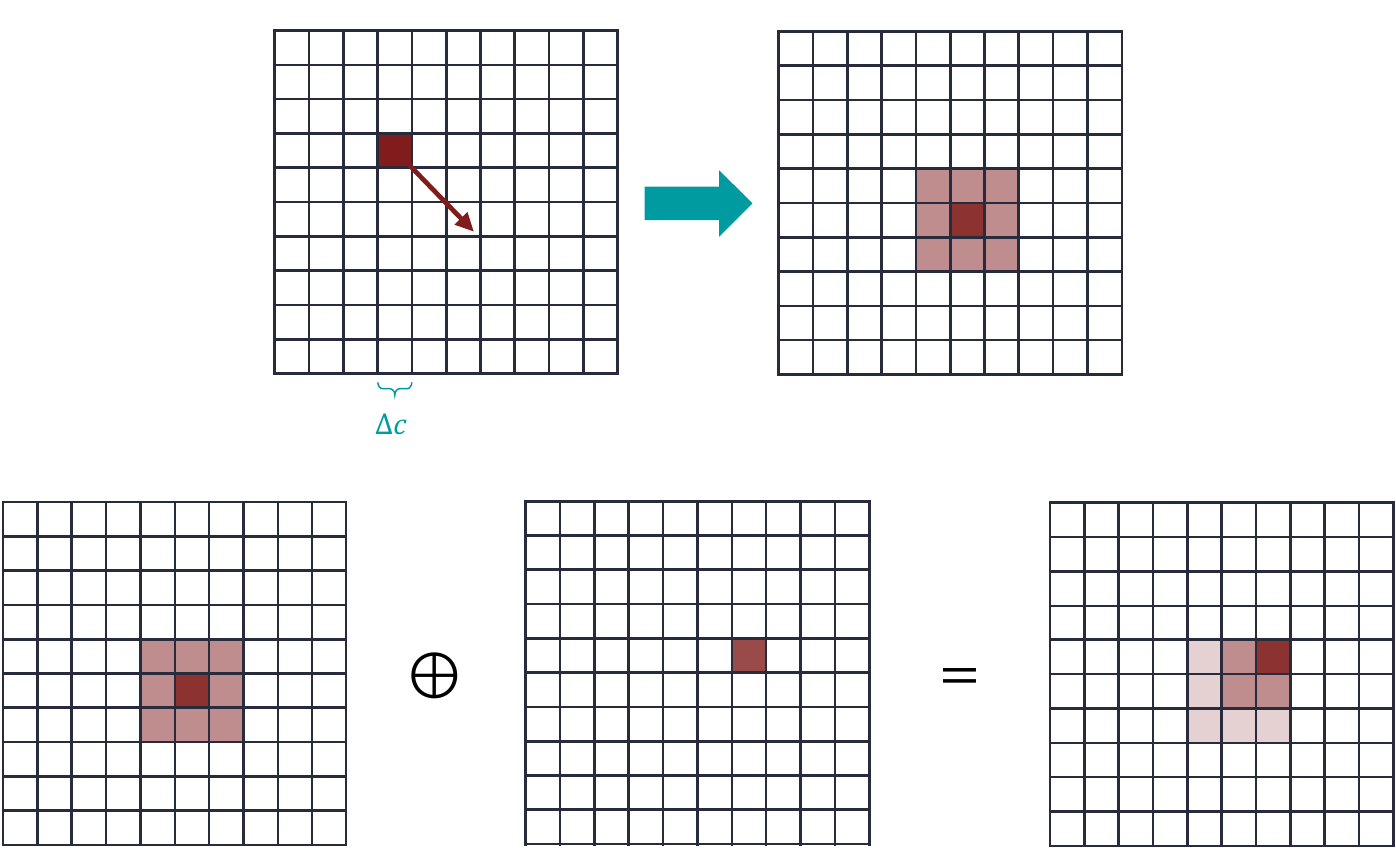}
	\caption{The occupancy state is predicted to its new location based on the velocity vector (red arrow) and distributed to adjacent cells according to the estimated state covariance. Then, the predicted state is fused with a new observation.}
	\label{fig:belief_prop_and_fuson}
\end{figure}

Fig. \ref{fig:belief_prop_and_fuson} shows a simple example where the occupancy state of a single cell is predicted and distributed according to the state covariance estimate. It is then fused with the most recent measurement. 

\section{Creation of Ground Truth DOGMas}

In order to assess the fusion algorithm introduced in this paper, we create ground truth $D^{(l_i)}$ in the simulation environment Virtual Test Drive (VTD) \cite{Neumann-Cosel2009}. They can be used as input to the introduced fusion process. We propose the following mechanism to create a ground truth $D^{(l_i)}$. Each vehicle is equipped with a virtual laser scanner that is placed at the same location as the laser scanner from which $D^{(l_i)}$ are computed. The laser scanner has an extremely high resolution and can provide the class of the object that produces a reflection. In this work, we assign each cell containing a reflection of the class road with $m_F = 1$ and $m_O = 0$ and all other cells containing a reflection with $m_F = 0$ and $m_O = 1$. Cells not containing any reflections are assigned with belief masses $m_F = 0$ and $m_O = 0$, except for cells that contain the ego-vehicle which are assigned with $m_F = 0$ and $m_O = 1$. The latter assignment is part of the self-reporting of connected vehicles which adds value if multiple $D^{(l_i)}$ are combined. The cell dynamics are taken from a ground truth object list provided by VTD. All cells containing a reflection produced by an object are assigned with the velocity of that object taken from the object list. The cells occupied by the ego-vehicle are also filled with the respective velocity data.

\begin{figure}
	\centering
	\includegraphics[trim=0 80 0 90,clip, width=3in]{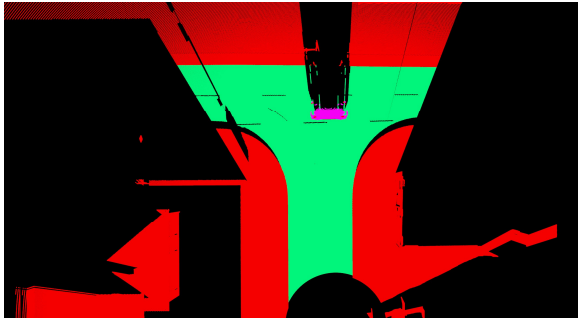}
	\caption{The labeled point cloud provided by an HD laser scanner in combination with a ground truth object list provided by VTD is used to create a ground truth DOGMa. In this image, we can see points associated with the class \textit{road surface} colored green, points associated with the class \textit{vehicle} are colored pink. All other points are colored red.}
	\label{fig:hd_lidar}
\end{figure}

In order to remove any errors introduced by the algorithm computing the $\hat{D}^{(l_i)}_{k+1|k}$, we only fuse ground truth $D^{(l_i)}$ in this work. This way, we can more easily trace back potential errors to the fusion process. Fig. \ref{fig:hd_lidar} shows an example of a ground truth point cloud from which a $D^{(l_i)}$ can be computed.

\section{Experimental Setup and Research Question} 

\begin{figure}[ht]
	\centering
	\includegraphics[trim=0 105 0 90,clip, width=3in]{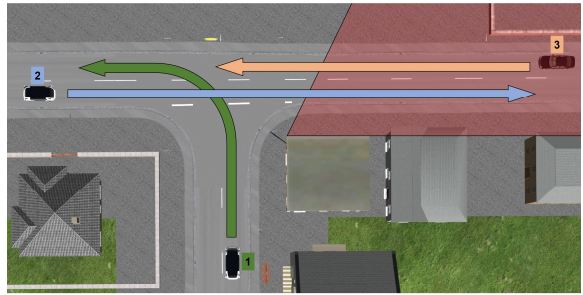}
	\caption{The connected vehicles ($i=1$) and ($i=2$) are each sending a lidar-based DOGMa $D^{(l_i)}_{k_i}$ to the CEM and receive $\hat{D}^{(l_1)}_{k_1+1|k_1}$ and $\hat{D}^{(l_2)}_{k_2+1|k_2}$ in return. Vehicle~$3$ moves slowly enough that vehicle~$1$ is able to make a left turn without waiting for vehicle~$3$. It is only admissible for vehicle~$1$ to make the left turn if it receives information on the occluded part colored red. This can happen through the CEM where the $D^{(l_i)}_{k_i}$ are fused.}
	\label{fig:scenario}
\end{figure}

We test the hypothesis that there is a reduction of uncertainty in $\hat{D}^{(l_i)}_{k+1|k}$ computed from a $\hat{D}^{(c)}_{k|k}$ compared to the $D^{(l_i)}_{k+1}$ computed in an individual vehicle. For the quantification of the uncertainty, we use the measures introduced in Section \ref{sec:measures_of_uncertainty}. Fig. \ref{fig:scenario} shows the analyzed scenario. Two connected vehicles are inside a traffic area and send their respective $D^{(l_i)}$ to the CEM. There is a third vehicle that is not connected to the CEM. It is occluded from the second vehicle's initial view which again is occluded from the first vehicle's initial view. We analyze the results with respect to the first vehicle ($i=1$). This vehicle can especially benefit from additional data. It takes a left turn, therefore potential traffic from the right has the right of way. A left turn with no relevant braking is only admissible if a sufficient reduction of uncertainty in occluded areas is possible. We do not assess what is considered sufficient. We only quantify the uncertainty. The scenario is set up such that it is possible for the first vehicle to make a left turn. The scenario starts at the moment where $D^{(l_1)}$ and $D^{(l_2)}$ first overlap. We also assess the mean belief mass $\overline{m}_F$. This measure indicates how the mean number of cells which are considered free develops throughout the scenario.

\section{Results and Discussion}

\begin{figure}[t]
	\begin{tikzpicture}
	\begin{axis}[smooth, enlarge x limits=false, ymin=0.0, ymax=1.0, 
	cycle list/RdGy-6,
	mark list fill={.!75!white},
	cycle multiindex* list={
		RdGy-6
		\nextlist
		my marks
		\nextlist
		[3 of]linestyles
		\nextlist
		very thick
		\nextlist
	},
	xlabel={time in seconds},
	x label style={at={(axis description cs:0.5,0.02)},anchor=north},
	legend entries={$\overline{H}\big(\hat{D}^{(l_1)}_{k_1+1|k_1}\big)$, $\overline{H}\big(D^{(l_1)}_{k_1+1}\big)$, $\overline{NS}\big(\hat{D}^{(l_1)}_{k_1+1|k_1}\big)$,  $\overline{NS}\big(D^{(l_1)}_{k_1+1}\big)$, $\overline{m}_F\big(\hat{D}^{(l_1)}_{k_1+1|k_1})\big)$, $\overline{m}_F\big(D^{(l_1)}_{k_1+1}\big)$},
	legend style={at={(0.5,-0.2)},anchor=north},
	legend cell align={left},
	legend columns=2,]
	\addplot+[mark repeat=30,mark phase=5, red!90!white, solid] coordinates 
	{( 0.0000 , 0.83 ) ( 0.1010 , 0.82 ) ( 0.2020 , 0.82 ) ( 0.3030 , 0.81 ) ( 0.4040 , 0.81 ) ( 0.5050 , 0.81 ) ( 0.6060 , 0.81 ) ( 0.7070 , 0.80 ) ( 0.8080 , 0.79 ) ( 0.9090 , 0.77 ) ( 1.0100 , 0.75 ) ( 1.1110 , 0.73 ) ( 1.2120 , 0.72 ) ( 1.3130 , 0.70 ) ( 1.4140 , 0.69 ) ( 1.5150 , 0.68 ) ( 1.6160 , 0.67 ) ( 1.7170 , 0.66 ) ( 1.8180 , 0.65 ) ( 1.9190 , 0.64 ) ( 2.0200 , 0.62 ) ( 2.1210 , 0.61 ) ( 2.2220 , 0.61 ) ( 2.3230 , 0.60 ) ( 2.4240 , 0.59 ) ( 2.5250 , 0.58 ) ( 2.6260 , 0.57 ) ( 2.7270 , 0.56 ) ( 2.8280 , 0.55 ) ( 2.9290 , 0.54 ) ( 3.0300 , 0.53 ) ( 3.1310 , 0.52 ) ( 3.2320 , 0.52 ) ( 3.3330 , 0.51 ) ( 3.4340 , 0.52 ) ( 3.5350 , 0.52 ) ( 3.6360 , 0.52 ) ( 3.7370 , 0.53 ) ( 3.8380 , 0.53 ) ( 3.9390 , 0.53 ) ( 4.0400 , 0.53 ) ( 4.1410 , 0.53 ) ( 4.2420 , 0.53 ) ( 4.3430 , 0.53 ) ( 4.4440 , 0.53 ) ( 4.5450 , 0.53 ) ( 4.6460 , 0.54 ) ( 4.7470 , 0.54 ) ( 4.8480 , 0.55 ) ( 4.9490 , 0.55 ) ( 5.0500 , 0.56 ) ( 5.1510 , 0.56 ) ( 5.2520 , 0.57 ) ( 5.3530 , 0.58 ) ( 5.4540 , 0.59 ) ( 5.5550 , 0.60 ) ( 5.6560 , 0.61 ) ( 5.7570 , 0.61 ) ( 5.8580 , 0.61 ) ( 5.9590 , 0.60 ) ( 6.0600 , 0.60 ) ( 6.1610 , 0.60 ) ( 6.2620 , 0.60 ) ( 6.3630 , 0.61 ) ( 6.4640 , 0.61 ) ( 6.5650 , 0.62 ) ( 6.6660 , 0.63 ) ( 6.7670 , 0.63 ) ( 6.8680 , 0.63 ) ( 6.9690 , 0.64 ) ( 7.0700 , 0.64 ) ( 7.1710 , 0.64 ) ( 7.2720 , 0.65 ) ( 7.3730 , 0.66 ) ( 7.4740 , 0.66 ) ( 7.5750 , 0.67 ) ( 7.6760 , 0.67 ) ( 7.7770 , 0.68 ) ( 7.8780 , 0.68 ) ( 7.9790 , 0.69 ) ( 8.0800 , 0.69 ) ( 8.1810 , 0.70 ) ( 8.2820 , 0.70 ) ( 8.3830 , 0.71 ) ( 8.4840 , 0.71 ) ( 8.5850 , 0.72 ) ( 8.6860 , 0.72 ) ( 8.7870 , 0.72 ) ( 8.8880 , 0.72 ) ( 8.9890 , 0.72 ) ( 9.0900 , 0.72 ) ( 9.1910 , 0.72 ) ( 9.2920 , 0.72 ) ( 9.3930 , 0.72 ) ( 9.4940 , 0.72 ) ( 9.5950 , 0.72 ) ( 9.6960 , 0.72 ) ( 9.7970 , 0.72 ) ( 9.8980 , 0.73 ) ( 9.9990 , 0.73 ) ( 10.1000 , 0.73 ) ( 10.2010 , 0.73 ) ( 10.3020 , 0.73 ) ( 10.4030 , 0.73 ) ( 10.5040 , 0.74 ) ( 10.6050 , 0.74 ) ( 10.7060 , 0.74 ) ( 10.8070 , 0.75 ) ( 10.9080 , 0.75 ) ( 11.0090 , 0.76 ) ( 11.1100 , 0.76 ) ( 11.2110 , 0.76 ) ( 11.3120 , 0.77 ) ( 11.4130 , 0.77 ) ( 11.5140 , 0.77 ) ( 11.6150 , 0.78 ) ( 11.7160 , 0.78 ) ( 11.8170 , 0.79 )};
	
	\addplot+[mark repeat=30,mark phase=5, red!50!white, dotted] coordinates 
	{( 0.0000 , 0.84 ) ( 0.1010 , 0.84 ) ( 0.2020 , 0.84 ) ( 0.3030 , 0.85 ) ( 0.4040 , 0.85 ) ( 0.5050 , 0.85 ) ( 0.6060 , 0.85 ) ( 0.7070 , 0.85 ) ( 0.8080 , 0.85 ) ( 0.9090 , 0.84 ) ( 1.0100 , 0.84 ) ( 1.1110 , 0.84 ) ( 1.2120 , 0.84 ) ( 1.3130 , 0.84 ) ( 1.4140 , 0.85 ) ( 1.5150 , 0.85 ) ( 1.6160 , 0.84 ) ( 1.7170 , 0.84 ) ( 1.8180 , 0.83 ) ( 1.9190 , 0.83 ) ( 2.0200 , 0.82 ) ( 2.1210 , 0.81 ) ( 2.2220 , 0.81 ) ( 2.3230 , 0.81 ) ( 2.4240 , 0.81 ) ( 2.5250 , 0.82 ) ( 2.6260 , 0.82 ) ( 2.7270 , 0.82 ) ( 2.8280 , 0.82 ) ( 2.9290 , 0.82 ) ( 3.0300 , 0.83 ) ( 3.1310 , 0.83 ) ( 3.2320 , 0.83 ) ( 3.3330 , 0.84 ) ( 3.4340 , 0.84 ) ( 3.5350 , 0.83 ) ( 3.6360 , 0.83 ) ( 3.7370 , 0.82 ) ( 3.8380 , 0.82 ) ( 3.9390 , 0.82 ) ( 4.0400 , 0.82 ) ( 4.1410 , 0.81 ) ( 4.2420 , 0.81 ) ( 4.3430 , 0.81 ) ( 4.4440 , 0.81 ) ( 4.5450 , 0.81 ) ( 4.6460 , 0.81 ) ( 4.7470 , 0.81 ) ( 4.8480 , 0.81 ) ( 4.9490 , 0.81 ) ( 5.0500 , 0.80 ) ( 5.1510 , 0.79 ) ( 5.2520 , 0.78 ) ( 5.3530 , 0.78 ) ( 5.4540 , 0.78 ) ( 5.5550 , 0.78 ) ( 5.6560 , 0.78 ) ( 5.7570 , 0.78 ) ( 5.8580 , 0.78 ) ( 5.9590 , 0.78 ) ( 6.0600 , 0.78 ) ( 6.1610 , 0.78 ) ( 6.2620 , 0.78 ) ( 6.3630 , 0.78 ) ( 6.4640 , 0.78 ) ( 6.5650 , 0.78 ) ( 6.6660 , 0.78 ) ( 6.7670 , 0.78 ) ( 6.8680 , 0.78 ) ( 6.9690 , 0.79 ) ( 7.0700 , 0.79 ) ( 7.1710 , 0.79 ) ( 7.2720 , 0.79 ) ( 7.3730 , 0.80 ) ( 7.4740 , 0.80 ) ( 7.5750 , 0.80 ) ( 7.6760 , 0.81 ) ( 7.7770 , 0.81 ) ( 7.8780 , 0.81 ) ( 7.9790 , 0.81 ) ( 8.0800 , 0.81 ) ( 8.1810 , 0.81 ) ( 8.2820 , 0.81 ) ( 8.3830 , 0.81 ) ( 8.4840 , 0.81 ) ( 8.5850 , 0.82 ) ( 8.6860 , 0.82 ) ( 8.7870 , 0.82 ) ( 8.8880 , 0.82 ) ( 8.9890 , 0.82 ) ( 9.0900 , 0.82 ) ( 9.1910 , 0.82 ) ( 9.2920 , 0.82 ) ( 9.3930 , 0.82 ) ( 9.4940 , 0.82 ) ( 9.5950 , 0.82 ) ( 9.6960 , 0.82 ) ( 9.7970 , 0.83 ) ( 9.8980 , 0.83 ) ( 9.9990 , 0.83 ) ( 10.1000 , 0.83 ) ( 10.2010 , 0.83 ) ( 10.3020 , 0.83 ) ( 10.4030 , 0.83 ) ( 10.5040 , 0.83 ) ( 10.6050 , 0.83 ) ( 10.7060 , 0.83 ) ( 10.8070 , 0.83 ) ( 10.9080 , 0.83 ) ( 11.0090 , 0.83 ) ( 11.1100 , 0.83 ) ( 11.2110 , 0.83 ) ( 11.3120 , 0.83 ) ( 11.4130 , 0.84 ) ( 11.5140 , 0.84 ) ( 11.6150 , 0.84 ) ( 11.7160 , 0.84 ) ( 11.8170 , 0.84 )};
	
	\addplot+[mark repeat=30,mark phase=10, blue!90!white, solid] coordinates 
	{( 0.0000 , 0.83 ) ( 0.1010 , 0.82 ) ( 0.2020 , 0.81 ) ( 0.3030 , 0.81 ) ( 0.4040 , 0.80 ) ( 0.5050 , 0.80 ) ( 0.6060 , 0.80 ) ( 0.7070 , 0.80 ) ( 0.8080 , 0.78 ) ( 0.9090 , 0.76 ) ( 1.0100 , 0.74 ) ( 1.1110 , 0.72 ) ( 1.2120 , 0.71 ) ( 1.3130 , 0.69 ) ( 1.4140 , 0.67 ) ( 1.5150 , 0.66 ) ( 1.6160 , 0.65 ) ( 1.7170 , 0.64 ) ( 1.8180 , 0.63 ) ( 1.9190 , 0.62 ) ( 2.0200 , 0.60 ) ( 2.1210 , 0.59 ) ( 2.2220 , 0.59 ) ( 2.3230 , 0.59 ) ( 2.4240 , 0.57 ) ( 2.5250 , 0.57 ) ( 2.6260 , 0.56 ) ( 2.7270 , 0.55 ) ( 2.8280 , 0.54 ) ( 2.9290 , 0.53 ) ( 3.0300 , 0.52 ) ( 3.1310 , 0.51 ) ( 3.2320 , 0.50 ) ( 3.3330 , 0.49 ) ( 3.4340 , 0.49 ) ( 3.5350 , 0.50 ) ( 3.6360 , 0.50 ) ( 3.7370 , 0.50 ) ( 3.8380 , 0.51 ) ( 3.9390 , 0.51 ) ( 4.0400 , 0.50 ) ( 4.1410 , 0.51 ) ( 4.2420 , 0.51 ) ( 4.3430 , 0.51 ) ( 4.4440 , 0.51 ) ( 4.5450 , 0.51 ) ( 4.6460 , 0.51 ) ( 4.7470 , 0.52 ) ( 4.8480 , 0.52 ) ( 4.9490 , 0.53 ) ( 5.0500 , 0.53 ) ( 5.1510 , 0.54 ) ( 5.2520 , 0.55 ) ( 5.3530 , 0.56 ) ( 5.4540 , 0.57 ) ( 5.5550 , 0.58 ) ( 5.6560 , 0.58 ) ( 5.7570 , 0.59 ) ( 5.8580 , 0.59 ) ( 5.9590 , 0.59 ) ( 6.0600 , 0.58 ) ( 6.1610 , 0.58 ) ( 6.2620 , 0.59 ) ( 6.3630 , 0.59 ) ( 6.4640 , 0.60 ) ( 6.5650 , 0.61 ) ( 6.6660 , 0.62 ) ( 6.7670 , 0.62 ) ( 6.8680 , 0.62 ) ( 6.9690 , 0.62 ) ( 7.0700 , 0.63 ) ( 7.1710 , 0.63 ) ( 7.2720 , 0.63 ) ( 7.3730 , 0.64 ) ( 7.4740 , 0.65 ) ( 7.5750 , 0.65 ) ( 7.6760 , 0.66 ) ( 7.7770 , 0.66 ) ( 7.8780 , 0.67 ) ( 7.9790 , 0.67 ) ( 8.0800 , 0.68 ) ( 8.1810 , 0.68 ) ( 8.2820 , 0.69 ) ( 8.3830 , 0.70 ) ( 8.4840 , 0.70 ) ( 8.5850 , 0.70 ) ( 8.6860 , 0.71 ) ( 8.7870 , 0.71 ) ( 8.8880 , 0.71 ) ( 8.9890 , 0.71 ) ( 9.0900 , 0.71 ) ( 9.1910 , 0.71 ) ( 9.2920 , 0.71 ) ( 9.3930 , 0.71 ) ( 9.4940 , 0.71 ) ( 9.5950 , 0.71 ) ( 9.6960 , 0.71 ) ( 9.7970 , 0.71 ) ( 9.8980 , 0.71 ) ( 9.9990 , 0.71 ) ( 10.1000 , 0.72 ) ( 10.2010 , 0.72 ) ( 10.3020 , 0.72 ) ( 10.4030 , 0.72 ) ( 10.5040 , 0.72 ) ( 10.6050 , 0.73 ) ( 10.7060 , 0.73 ) ( 10.8070 , 0.74 ) ( 10.9080 , 0.74 ) ( 11.0090 , 0.75 ) ( 11.1100 , 0.75 ) ( 11.2110 , 0.75 ) ( 11.3120 , 0.76 ) ( 11.4130 , 0.76 ) ( 11.5140 , 0.77 ) ( 11.6150 , 0.77 ) ( 11.7160 , 0.77 ) ( 11.8170 , 0.78 )};
	
	\addplot+[mark repeat=30,mark phase=10, blue!50!white, dotted] coordinates 
	{( 0.0000 , 0.840 ) ( 0.1010 , 0.841 ) ( 0.2020 , 0.844 ) ( 0.3030 , 0.846 ) ( 0.4040 , 0.847 ) ( 0.5050 , 0.848 ) ( 0.6060 , 0.847 ) ( 0.7070 , 0.848 ) ( 0.8080 , 0.845 ) ( 0.9090 , 0.842 ) ( 1.0100 , 0.843 ) ( 1.1110 , 0.844 ) ( 1.2120 , 0.843 ) ( 1.3130 , 0.845 ) ( 1.4140 , 0.847 ) ( 1.5150 , 0.846 ) ( 1.6160 , 0.842 ) ( 1.7170 , 0.838 ) ( 1.8180 , 0.832 ) ( 1.9190 , 0.826 ) ( 2.0200 , 0.818 ) ( 2.1210 , 0.813 ) ( 2.2220 , 0.811 ) ( 2.3230 , 0.812 ) ( 2.4240 , 0.815 ) ( 2.5250 , 0.819 ) ( 2.6260 , 0.821 ) ( 2.7270 , 0.821 ) ( 2.8280 , 0.823 ) ( 2.9290 , 0.825 ) ( 3.0300 , 0.828 ) ( 3.1310 , 0.831 ) ( 3.2320 , 0.835 ) ( 3.3330 , 0.842 ) ( 3.4340 , 0.838 ) ( 3.5350 , 0.834 ) ( 3.6360 , 0.829 ) ( 3.7370 , 0.825 ) ( 3.8380 , 0.825 ) ( 3.9390 , 0.823 ) ( 4.0400 , 0.816 ) ( 4.1410 , 0.814 ) ( 4.2420 , 0.815 ) ( 4.3430 , 0.815 ) ( 4.4440 , 0.815 ) ( 4.5450 , 0.812 ) ( 4.6460 , 0.810 ) ( 4.7470 , 0.813 ) ( 4.8480 , 0.814 ) ( 4.9490 , 0.810 ) ( 5.0500 , 0.803 ) ( 5.1510 , 0.794 ) ( 5.2520 , 0.784 ) ( 5.3530 , 0.779 ) ( 5.4540 , 0.778 ) ( 5.5550 , 0.778 ) ( 5.6560 , 0.777 ) ( 5.7570 , 0.777 ) ( 5.8580 , 0.777 ) ( 5.9590 , 0.778 ) ( 6.0600 , 0.780 ) ( 6.1610 , 0.779 ) ( 6.2620 , 0.778 ) ( 6.3630 , 0.777 ) ( 6.4640 , 0.776 ) ( 6.5650 , 0.777 ) ( 6.6660 , 0.779 ) ( 6.7670 , 0.781 ) ( 6.8680 , 0.784 ) ( 6.9690 , 0.787 ) ( 7.0700 , 0.789 ) ( 7.1710 , 0.792 ) ( 7.2720 , 0.795 ) ( 7.3730 , 0.798 ) ( 7.4740 , 0.801 ) ( 7.5750 , 0.804 ) ( 7.6760 , 0.806 ) ( 7.7770 , 0.807 ) ( 7.8780 , 0.809 ) ( 7.9790 , 0.810 ) ( 8.0800 , 0.811 ) ( 8.1810 , 0.812 ) ( 8.2820 , 0.812 ) ( 8.3830 , 0.813 ) ( 8.4840 , 0.814 ) ( 8.5850 , 0.815 ) ( 8.6860 , 0.816 ) ( 8.7870 , 0.817 ) ( 8.8880 , 0.817 ) ( 8.9890 , 0.816 ) ( 9.0900 , 0.815 ) ( 9.1910 , 0.816 ) ( 9.2920 , 0.818 ) ( 9.3930 , 0.818 ) ( 9.4940 , 0.820 ) ( 9.5950 , 0.822 ) ( 9.6960 , 0.824 ) ( 9.7970 , 0.825 ) ( 9.8980 , 0.827 ) ( 9.9990 , 0.828 ) ( 10.1000 , 0.827 ) ( 10.2010 , 0.827 ) ( 10.3020 , 0.826 ) ( 10.4030 , 0.826 ) ( 10.5040 , 0.826 ) ( 10.6050 , 0.827 ) ( 10.7060 , 0.827 ) ( 10.8070 , 0.828 ) ( 10.9080 , 0.830 ) ( 11.0090 , 0.832 ) ( 11.1100 , 0.833 ) ( 11.2110 , 0.834 ) ( 11.3120 , 0.835 ) ( 11.4130 , 0.835 ) ( 11.5140 , 0.836 ) ( 11.6150 , 0.836 ) ( 11.7160 , 0.836 ) ( 11.8170 , 0.836 )};
	
	\addplot+[mark repeat=30,mark phase=5, green!90!white!60!black, solid ] coordinates 
	{( 0.0000 , 0.044 ) ( 0.1010 , 0.045 ) ( 0.2020 , 0.046 ) ( 0.3030 , 0.048 ) ( 0.4040 , 0.049 ) ( 0.5050 , 0.051 ) ( 0.6060 , 0.052 ) ( 0.7070 , 0.052 ) ( 0.8080 , 0.054 ) ( 0.9090 , 0.055 ) ( 1.0100 , 0.058 ) ( 1.1110 , 0.061 ) ( 1.2120 , 0.065 ) ( 1.3130 , 0.069 ) ( 1.4140 , 0.073 ) ( 1.5150 , 0.080 ) ( 1.6160 , 0.087 ) ( 1.7170 , 0.095 ) ( 1.8180 , 0.103 ) ( 1.9190 , 0.112 ) ( 2.0200 , 0.122 ) ( 2.1210 , 0.132 ) ( 2.2220 , 0.134 ) ( 2.3230 , 0.137 ) ( 2.4240 , 0.139 ) ( 2.5250 , 0.141 ) ( 2.6260 , 0.142 ) ( 2.7270 , 0.143 ) ( 2.8280 , 0.144 ) ( 2.9290 , 0.145 ) ( 3.0300 , 0.146 ) ( 3.1310 , 0.148 ) ( 3.2320 , 0.149 ) ( 3.3330 , 0.151 ) ( 3.4340 , 0.153 ) ( 3.5350 , 0.157 ) ( 3.6360 , 0.160 ) ( 3.7370 , 0.163 ) ( 3.8380 , 0.167 ) ( 3.9390 , 0.169 ) ( 4.0400 , 0.171 ) ( 4.1410 , 0.173 ) ( 4.2420 , 0.174 ) ( 4.3430 , 0.175 ) ( 4.4440 , 0.176 ) ( 4.5450 , 0.177 ) ( 4.6460 , 0.176 ) ( 4.7470 , 0.175 ) ( 4.8480 , 0.174 ) ( 4.9490 , 0.174 ) ( 5.0500 , 0.173 ) ( 5.1510 , 0.172 ) ( 5.2520 , 0.170 ) ( 5.3530 , 0.168 ) ( 5.4540 , 0.166 ) ( 5.5550 , 0.165 ) ( 5.6560 , 0.164 ) ( 5.7570 , 0.164 ) ( 5.8580 , 0.164 ) ( 5.9590 , 0.164 ) ( 6.0600 , 0.168 ) ( 6.1610 , 0.170 ) ( 6.2620 , 0.167 ) ( 6.3630 , 0.157 ) ( 6.4640 , 0.144 ) ( 6.5650 , 0.134 ) ( 6.6660 , 0.125 ) ( 6.7670 , 0.120 ) ( 6.8680 , 0.116 ) ( 6.9690 , 0.113 ) ( 7.0700 , 0.109 ) ( 7.1710 , 0.106 ) ( 7.2720 , 0.104 ) ( 7.3730 , 0.102 ) ( 7.4740 , 0.101 ) ( 7.5750 , 0.098 ) ( 7.6760 , 0.096 ) ( 7.7770 , 0.095 ) ( 7.8780 , 0.093 ) ( 7.9790 , 0.092 ) ( 8.0800 , 0.092 ) ( 8.1810 , 0.092 ) ( 8.2820 , 0.091 ) ( 8.3830 , 0.091 ) ( 8.4840 , 0.090 ) ( 8.5850 , 0.090 ) ( 8.6860 , 0.089 ) ( 8.7870 , 0.088 ) ( 8.8880 , 0.087 ) ( 8.9890 , 0.086 ) ( 9.0900 , 0.085 ) ( 9.1910 , 0.085 ) ( 9.2920 , 0.084 ) ( 9.3930 , 0.083 ) ( 9.4940 , 0.082 ) ( 9.5950 , 0.081 ) ( 9.6960 , 0.080 ) ( 9.7970 , 0.079 ) ( 9.8980 , 0.078 ) ( 9.9990 , 0.077 ) ( 10.1000 , 0.077 ) ( 10.2010 , 0.076 ) ( 10.3020 , 0.075 ) ( 10.4030 , 0.074 ) ( 10.5040 , 0.073 ) ( 10.6050 , 0.071 ) ( 10.7060 , 0.070 ) ( 10.8070 , 0.069 ) ( 10.9080 , 0.067 ) ( 11.0090 , 0.066 ) ( 11.1100 , 0.065 ) ( 11.2110 , 0.064 ) ( 11.3120 , 0.063 ) ( 11.4130 , 0.063 ) ( 11.5140 , 0.062 ) ( 11.6150 , 0.061 ) ( 11.7160 , 0.060 ) ( 11.8170 , 0.059 )};
	
	\addplot+[mark repeat=30,mark phase=10, green!40!white!70!black, dotted] coordinates 
	{( 0.0000 , 0.053 ) ( 0.1010 , 0.053 ) ( 0.2020 , 0.053 ) ( 0.3030 , 0.053 ) ( 0.4040 , 0.053 ) ( 0.5050 , 0.053 ) ( 0.6060 , 0.053 ) ( 0.7070 , 0.053 ) ( 0.8080 , 0.053 ) ( 0.9090 , 0.054 ) ( 1.0100 , 0.054 ) ( 1.1110 , 0.054 ) ( 1.2120 , 0.054 ) ( 1.3130 , 0.054 ) ( 1.4140 , 0.053 ) ( 1.5150 , 0.053 ) ( 1.6160 , 0.053 ) ( 1.7170 , 0.053 ) ( 1.8180 , 0.053 ) ( 1.9190 , 0.054 ) ( 2.0200 , 0.054 ) ( 2.1210 , 0.055 ) ( 2.2220 , 0.055 ) ( 2.3230 , 0.056 ) ( 2.4240 , 0.057 ) ( 2.5250 , 0.058 ) ( 2.6260 , 0.060 ) ( 2.7270 , 0.061 ) ( 2.8280 , 0.063 ) ( 2.9290 , 0.064 ) ( 3.0300 , 0.064 ) ( 3.1310 , 0.063 ) ( 3.2320 , 0.062 ) ( 3.3330 , 0.061 ) ( 3.4340 , 0.061 ) ( 3.5350 , 0.061 ) ( 3.6360 , 0.061 ) ( 3.7370 , 0.060 ) ( 3.8380 , 0.060 ) ( 3.9390 , 0.060 ) ( 4.0400 , 0.060 ) ( 4.1410 , 0.060 ) ( 4.2420 , 0.060 ) ( 4.3430 , 0.062 ) ( 4.4440 , 0.064 ) ( 4.5450 , 0.066 ) ( 4.6460 , 0.069 ) ( 4.7470 , 0.071 ) ( 4.8480 , 0.071 ) ( 4.9490 , 0.072 ) ( 5.0500 , 0.073 ) ( 5.1510 , 0.077 ) ( 5.2520 , 0.080 ) ( 5.3530 , 0.080 ) ( 5.4540 , 0.081 ) ( 5.5550 , 0.081 ) ( 5.6560 , 0.081 ) ( 5.7570 , 0.081 ) ( 5.8580 , 0.080 ) ( 5.9590 , 0.080 ) ( 6.0600 , 0.079 ) ( 6.1610 , 0.078 ) ( 6.2620 , 0.077 ) ( 6.3630 , 0.077 ) ( 6.4640 , 0.076 ) ( 6.5650 , 0.076 ) ( 6.6660 , 0.075 ) ( 6.7670 , 0.074 ) ( 6.8680 , 0.073 ) ( 6.9690 , 0.073 ) ( 7.0700 , 0.074 ) ( 7.1710 , 0.072 ) ( 7.2720 , 0.070 ) ( 7.3730 , 0.068 ) ( 7.4740 , 0.066 ) ( 7.5750 , 0.064 ) ( 7.6760 , 0.062 ) ( 7.7770 , 0.062 ) ( 7.8780 , 0.061 ) ( 7.9790 , 0.061 ) ( 8.0800 , 0.061 ) ( 8.1810 , 0.061 ) ( 8.2820 , 0.061 ) ( 8.3830 , 0.061 ) ( 8.4840 , 0.061 ) ( 8.5850 , 0.060 ) ( 8.6860 , 0.060 ) ( 8.7870 , 0.060 ) ( 8.8880 , 0.059 ) ( 8.9890 , 0.058 ) ( 9.0900 , 0.057 ) ( 9.1910 , 0.055 ) ( 9.2920 , 0.054 ) ( 9.3930 , 0.053 ) ( 9.4940 , 0.052 ) ( 9.5950 , 0.051 ) ( 9.6960 , 0.051 ) ( 9.7970 , 0.050 ) ( 9.8980 , 0.050 ) ( 9.9990 , 0.050 ) ( 10.1000 , 0.051 ) ( 10.2010 , 0.051 ) ( 10.3020 , 0.051 ) ( 10.4030 , 0.052 ) ( 10.5040 , 0.052 ) ( 10.6050 , 0.052 ) ( 10.7060 , 0.053 ) ( 10.8070 , 0.053 ) ( 10.9080 , 0.054 ) ( 11.0090 , 0.054 ) ( 11.1100 , 0.055 ) ( 11.2110 , 0.055 ) ( 11.3120 , 0.055 ) ( 11.4130 , 0.056 ) ( 11.5140 , 0.056 ) ( 11.6150 , 0.057 ) ( 11.7160 , 0.057 ) ( 11.8170 , 0.058 )};
	
	\end{axis}
	
	\end{tikzpicture}
	\caption{The plots show how the mean Shannon entropy $\overline{H}$, the mean non-specificity measure $\overline{NS}$ and the mean belief mass for free cells $\overline{m}_F$ develop throughout the analyzed scenario. We apply the measures to $\hat{D}^{(l_1)}_{k_1+1|k_1}$, which is provided as a service by the CEM, and to $D^{(l_1)}_{k_1}$, which is computed by vehicle~$1$.}
	\label{fig:results}
\end{figure}
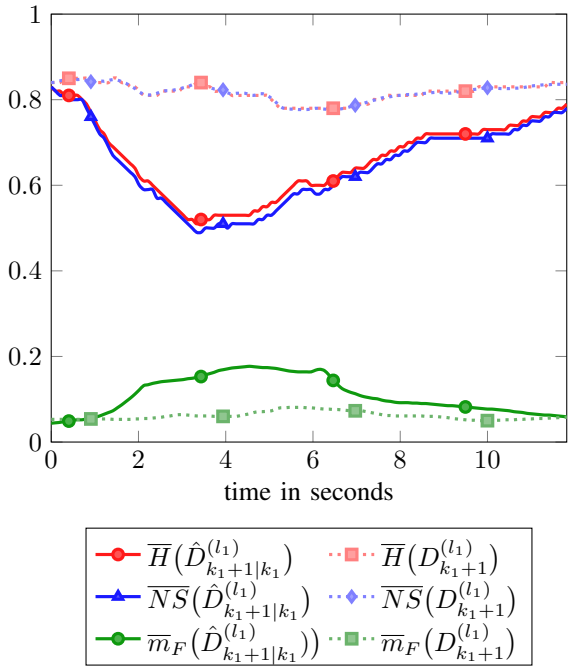

The results depicted in Fig. \ref{fig:results} confirm the stated hypothesis. Both measures of uncertainty $\overline{H}$ and $\overline{NS}$ are strictly smaller for $\hat{D}^{(l_1)}_{k_1+1|k_1}$ than for $D^{(l_1)}_{k_1+1}$, except for the start of the scenario where they are approximately equal. The latter is expected since, at the start of the scenario, there is no overlap in the $D^{(l_i)}$, thus no additional information can be contained in any $\hat{D}^{(l_i)}$. The Shannon entropy and the non-specificity measure develop very similarly throughout the scenario. This is expected since only very little conflict exists between the fused DOGMas. The similarity acts as an indicator that the synchronization and fusion process works correctly. Misaligned data would lead to conflicting information which would be captured by $\overline{H}$ but not by $\overline{NS}$.
The decrease in uncertainty while the overlapping area increases is steeper than the increase of uncertainty while the overlapping area decreases. This reflects the persistence of information in $D^{(c)}$. Even at times when an area is not directly visible to any vehicle, uncertainty in $\hat{D}^{(l_i)}$ may be decreased by the incorporation of past data. This does not strictly imply that a state is sufficiently certain for a planned maneuver through these areas. 
The mean belief mass $\overline{m}_F$ is strictly larger for $\hat{D}^{(l_1)}_{k_1+1|k_1}$ than for $D^{(l_1)}_{k_1+1}$, except for the start and the end of the scenario. This is important because a reduction of uncertainty only for occupied cells would not give us as many possibilities to plan maneuvers. Cells associated with a high belief mass $m_F$ are necessary to plan trajectories.

\begin{figure}[ht]
	\centering
	\includegraphics[angle=-90,origin=c, trim=20 0 38 0,clip, width=3in]{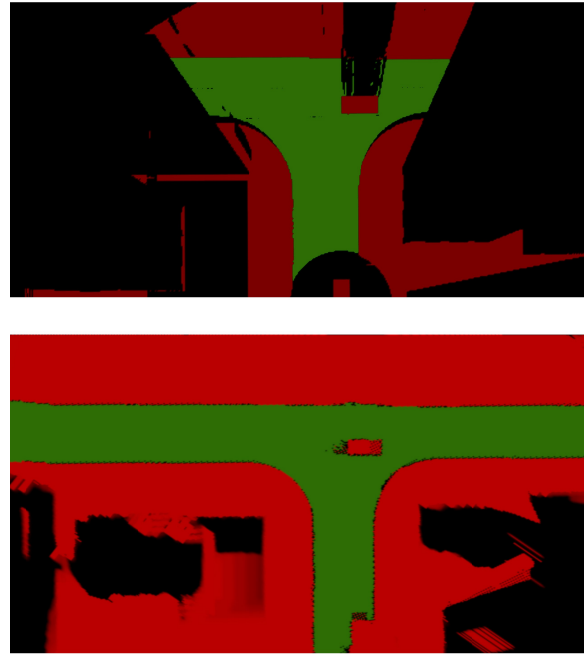}
	\caption{The upper image depicts the belief masses $m_F$ (green) and the belief masses $m_O$ (red) of $D^{(l_1)}_{k_1}$ which includes the self-reported pose and dimensions of vehicle~$1$. The bottom image shows $m_F$ and $m_o$ of $\hat{D}^{(l_1)}_{k_1+1|k_1}$ in the same colors.}
	\label{fig:qualitative_result}
\end{figure}

The qualitative results depicted in Fig. \ref{fig:qualitative_result} show how $\hat{D}^{(l_1)}_{k_1+1|k_1}$ contains substantially more cells associated with a large belief mass $m_F$ or $m_O$ than $D^{(l_1)}_{k_1+1}$. By having access to the DOGMa provided by the CEM, vehicle~$1$ has information on the cells that are occluded in its direct view. Since vehicle~$3$ is not in the vicinity of the T junction yet, the left turn may now be considered admissible.

\section{Conclusion}

In our research, we have developed a cloud-based mechanism as part of the CEM, which is capable of fusing multiple non-synchronized evidential DOGMas. The mechanism incorporates the self-reported poses and dimensions of connected vehicles. It also takes into account the current latency between vehicles and the cloud to predict the state of a fused DOGMa at the time it arrives in a vehicle. We show that the uncertainty in the DOGMas provided by the CEM can be reduced substantially compared to DOGMas computed in the individual vehicles. The results motivate further research because the decrease in uncertainty may in the future allow behavior of automated and connected vehicles that is more safe, more efficient and more convenient. 
There are two directions in which more research is especially required. First of all, the developed mechanism shall be tested and optimized for a real-world use case. Due to restrictions in bandwidth, latency, processing power of current technology and a lack of efficiency in the current implementation of the mechanism, the latter is not yet viable for usage in a real-world setting. Second, the input to the fusion mechanism presumably is a lot more noisy coming from real vehicles. This noise is not yet sufficiently incorporated in the fusion process. Future implementations will need to take it into account. 
The smaller the rate at which data arrives in the CEM and the larger the latency between vehicles, the less accurate the currently used motion model is. Since the CEM is capable of recording a lot of data over time, a learned motion model seems promising. This and other extensions to the described mechanism will be part of future research.

\addtolength{\textheight}{-12cm}   





\begin{thebibliography}{99}

\bibitem{Woopen2018} T. Woopen et al., "{UNICAR}agil - {D}isruptive {M}odular {A}rchitectures for {A}gile, {A}utomated {V}ehicle {C}oncepts", 27th Aachen Colloquium Automobile and Engine Technology 2018, Aachen, 2018
\bibitem{Kampmann2019} A. Kampmann et al., "A Dynamic Service-Oriented Software Architecture for Highly Automated Vehicles," 2019 IEEE Intelligent Transportation Systems Conference (ITSC), Auckland, New Zealand, 2019, pp. 2101-2108. doi: 10.1109/ITSC.2019.8916841
\bibitem{Lampe2019} B. Lampe, T. Woopen, and Lutz Eckstein, "Collective Driving - Cloud Services for Automated Vehicles in UNICARagil", 28th Aachen Colloquium Automobile and Engine Technology 2019, Aachen, 2019. doi: 10.18154/RWTH-2019-10061
\bibitem{Winner2015} H. Winner et al., Handbuch  Fahrerassistenzsysteme:  Grundlagen,  Komponenten und Systeme f\"ur aktive Sicherheit und Komfort, 3. \"uberarbeitete und erg\"anzte Auflage, ATZ/MTZFachbuch, Wiesbaden: Springer Vieweg, 2015
\bibitem{Thrun2006} S. Thrun, W. Burgard, and D. Fox, "Probabilistic robotics ( Intelligent robotics and autonomous agents)", Cambridge, Massachusetts, The MIT Press, 2006
\bibitem{Camarda2018} F. Camarda, F. Davoine and V. Cherfaoui, "Fusion of evidential occupancy grids for cooperative perception," 13th Annual Conference on System of Systems Engineering (SoSE), Paris, 2018, pp. 284-290. doi: 10.1109/SYSOSE.2018.8428723
\bibitem{Dezert2015} J. Dezert, J. Moras and B. Pannetier, "Environment perception using grid occupancy estimation with belief functions," 2015 18th International Conference on Information Fusion (Fusion), Washington, DC,  pp. 1070-1077. 2015
\bibitem{Grimmer2017} A. Grimmer, J. Clemens, and R. Wille, "Formal methods for reasoning and uncertainty reduction in evidential grid maps", International Journal of Approximate Reasoning, vol. 87, pp. 23-39, Aug 2017
\bibitem{Elfes1989} A. Elfes, "Using occupancy grids for mobile robot perception and navigation," Computer, vol. 22, no. 6, pp. 46-57, 1989.
\bibitem{Nuss2016} D. Nuss et al., "A Random Finite Set Approach for Dynamic Occupancy Grid Maps with Real-Time Application", The International Journal of Robotics Research, 2016. doi: 10.1177/0278364918775523
\bibitem{Shafer1976} G. Shafer. "A Mathematical Theory of Evidence", Princeton University Press, 1976
\bibitem{Dubois1985}  D. Dubois, H. Prade, "A note on measures of specificity for fuzzy sets", International Journal of General Systems, 1985, pp. 279-283. DOI: 10.1080/03081078508934893 
\bibitem{Reineking2014} T. Reineking,  J. Clemens, "Dimensions of Uncertainty in Evidential Grid Maps", Spatial Cognition IX: International Conference, Proceedings pp.283-298, Bremen, Germany, Sep 2014
\bibitem{Neumann-Cosel2009} K. von Neumann-Cosel, M. Dupuis, C. Weiss, "Virtual Test Drive-Provision of a Consistent Tool-Set for [D HS V]-in-the-Loop", Proceedings on Driving Simulation Conference Europe, 2009.

\end{thebibliography}
\end{document}